\documentclass[pra,twocolumn,showpacs]{revtex4}
\usepackage{times}
\usepackage{amsbsy,amssymb}
\usepackage{graphicx,color}

\def\one{{\mathchoice {\rm 1\mskip-4mu l} {\rm 1\mskip-4mu l} {\rm
1\mskip-4.5mu l} {\rm 1\mskip-5mu l}}}
\newcommand{\ket}[1]{|{#1}\rangle}
\newcommand{\bra}[1]{\langle{#1}|}

\providecommand{\ignore}[1]{}

\def\openone{\leavevmode\hbox{\small1\kern-3.8pt\normalsize1}}
\def\RR{{\rm I\kern-.2emR}}

\def\fu{\mathfrak{u}}

\def\fh{\mathfrak{h}}

\def\fsu{\mathfrak{su}}

\def\fso{\mathfrak{so}}

\def\openone{\leavevmode\hbox{\small1\kern-3.8pt\normalsize1}}
\def\RR{{\rm I\kern-.2emR}}

\def\hh{\hat{h}}

\def\hO{\hat{O}}
\def\hA{\hat{A}}

\def\hQ{\hat{Q}}
\def\he{\hat{e}}

\def\hg{\hat{g}}

\providecommand{\ignore}[1]{}

\newcommand{\shortqph}[1]{}

\begin{document}


\title{Lower bounds for the fidelity of entangled state preparation}


\author{Rolando D. Somma}
\email[]{somma@lanl.gov}
\affiliation{Los Alamos National Laboratory, MS D454, Los Alamos, NM 87545}
\author{John Chiaverini}
\affiliation{Los Alamos National Laboratory, MS D454, Los Alamos, NM 87545}
\author{Dana J. Berkeland}
\affiliation{Los Alamos National Laboratory, MS D454, Los Alamos, NM 87545}


\date{\today}
\begin{abstract}
Estimating the fidelity of state preparation in multi-qubit systems
is generally a time-consuming task.  Nevertheless, this complexity can be reduced if the desired state can be characterized by certain symmetries measurable with the corresponding experimental setup.  In this paper we give simple expressions to estimate the fidelity  of multi-qubit state preparation for rotational-invariant, stabilizer, and generalized coherent states.  We specifically discuss the cat, W-type, and generalized coherent states, and obtain efficiently measurable lower bounds for the fidelity.  We use these techniques to estimate the fidelity of a quantum simulation of an Ising-like interacting model using two trapped ions.  These results are directly applicable to experiments using fidelity-based entanglement witnesses, such as quantum simulations and quantum computation.
\end{abstract}



\maketitle

\section{Introduction}
\label{intro}

Highly entangled states provide required resources for quantum information processing (QIP), a developing field advancing both the fundamental understanding of quantum systems and novel technologies.  Entangled states are used to encode qubits for fault-tolerant quantum computation~\cite{nielsen00} and for more efficient quantum state readout~\cite{schaetz05}.  Entangled states are used for quantum communication over long distances and teleportation protocols~\cite{ben93}.  Finally, highly entangled states are central to many-body quantum simulations, whose power lies in their ability to coherently manipulate such states for later analysis~\cite{lloy96,ort01,som02,por04}.  Entangled-state preparation in any QIP system, and its verification, is thus of paramount importance.

One successful architecture for QIP is the trapped-ion system, in which qubits are encoded in the internal electronic states of ions, and laser fields can control the collective internal and external states of the ions.  Recently~\cite{leibfried05,haffner05}, multi-qubit entanglement has been experimentally demonstrated in these devices.  In Ref.~\cite{haffner05}, quantum state tomography (QST)~\cite{james01,thew02} was employed to verify that W-type states for up to $N=8$ ions (qubits) were produced.  Since the dimension of the Hilbert space ${\cal H}$ associated with a quantum system increases exponentially with the system size (as does the dimension of the density matrix), performing full QST is, in general, extremely inefficient for large systems.  For example, realizing QST on an ion-trap device requires on the order of ${\cal O}(3^N)$ measurements, where $N$ is the number of qubits involved that are measured in the $x$, $y$, and $z$-bases.  In~\cite{haffner05} the full QST process for $N=8$ ions required $656,100$ measurements over ten hours.  This extremely large data set reduced errors due to quantum projection noise~\cite{itano93}, until other sources of error (such as imperfect optical pumping, ion addressing errors, non-resonant excitations and optical decoherence) dominate.  Such examples illustrate a potential roadblock to practical implementation of large-scale QIP:  it is impossible to exploit the speedups associated with QIP if an exponentially-large amount of processing must be performed to verify the creation of the desired states.

It is important then to investigate efficient methods to estimate the reliability of experimental quantum state preparation.  Here we point out that many useful entangled states have certain symmetries which allow fidelity determination without full QST.   For these states, an efficient number (polynomial in $N$) of measurements is sufficient to obtain lower bounds for the fidelity.  A similar technique has been used to determine a lower bound on the fidelity of several-particle cat states~\cite{leibfried05}; we describe and generalize such methods.  To see this, we use the {\em quantum fidelity} as a measure of  the {\em distance} between quantum states~\cite{nielsen00}. Specifically, the quantum fidelity ${\cal F}$  between the actual state prepared in the laboratory  $\rho_l$, which is in general mixed (i.e., ${\sf Tr}(\rho_l^2)<1$), and the desired pure state $\ket{\psi}$ to be prepared is defined by
\begin{equation}
\label{fid1}
{\cal F}(\rho_l,\rho_\psi) = \sqrt{\bra {\psi} \rho_l \ket{\psi} }= [{\sf Tr} (\rho_l \rho_\psi)]^{1/2}.
\end{equation}
Equation~(\ref{fid1}) can be evaluated by measuring the expectation value of the density operator $\rho_\psi= \ket{\psi} \bra{\psi}$ over the state $\rho_l$. For example, if $\ket{\psi}$ is a product state, then $\rho_\psi$ has only one non-zero matrix element (in the right basis) that is along its diagonal.  The fidelity ${\cal F}(\rho_l,\rho_\psi)$ can be simply obtained by repeatedly preparing $\rho_l$ and then measuring the population of the state $\ket{\psi}$.

More generally, the density matrix of an $N$-qubit system is a linear combination of operators belonging to the $\fu(2^N)$ algebra:
\begin{equation}
\label{dens-decomp}
\rho = \sum_{\alpha_1,\cdots,\alpha_N} c^\rho_{\alpha_1,\cdots,\alpha_N}
 (\sigma_{\alpha_1}^1   \otimes \cdots \otimes
\sigma_{\alpha_N}^N),
\end{equation}
where the subscripts $\alpha_j=0,1,2,3$ correspond to the Pauli operators $\one$, $\sigma_x$, $\sigma_y$, and $\sigma_z$, respectively. (The symbol $\otimes$ represents the matrix tensor product.) These operators are given by
\begin{eqnarray}
\label{pauli}
\one&=&
\pmatrix{ 1 & 0 \cr 0 & 1
} , \
\sigma_x=
\pmatrix{ 0 & 1 \cr 1 & 0
} , \\
\nonumber
\sigma_y&=&
\pmatrix{ 0 & -i \cr i & 0
} , \
\sigma_z=
\pmatrix{ 1 & 0 \cr 0 & -1
}.
\end{eqnarray}
In particular,
 $\sigma_{\alpha_j}^j= \one^1 \otimes \cdots \otimes \one^{j-1} \otimes \sigma^j_{\alpha_j}\otimes \one^{j+1} \cdots \otimes \one^N$, with
the Pauli matrix $\sigma_{\alpha_j}$ being located at the $j$th position in the decomposition. From now on, we remove the symbol $\otimes$ from the products of Pauli operators.  We also adopt the convention $\langle \hA \rangle_{\rho_l}\equiv {\sf Tr}[\rho_l \hA]$.

The real coefficients  $c^\rho_{\alpha_1,\alpha_2,\cdots,\alpha_N} $ are given by
\begin{eqnarray}
c^\rho_{0,\cdots,0}&=&2^{-N},  \\
\nonumber
c^\rho_{\alpha_1,\cdots,\alpha_N} &=&2^{-N} {\sf Tr} [\rho(
\sigma_{\alpha_1}^1 \cdots  \sigma_{\alpha_N}^N ) ] \ \text{(otherwise)}.
\end{eqnarray}
Then,
\begin{equation}
\label{fid2}
{\cal F}^2(\rho_l,\rho_\psi)= \sum_{\alpha_1,\cdots,\alpha_N} c^{\rho_l}_{\alpha_1,\cdots,\alpha_N}
c^{\rho_\psi}_{\alpha_1,\cdots,\alpha_N},
\end{equation}
and full QST is generally needed to estimate the coefficients $c^{\rho_l}_{\alpha_1,\cdots,\alpha_N}$ required
to evaluate Eq.~(\ref{fid2}).
However, if the state $\ket{\psi}$ can be uniquely characterized by certain symmetries, some of the coefficients $c^{\rho_\psi}_{\alpha_1,\cdots,\alpha_N}$ will vanish and the corresponding
$c^{\rho_l}_{\alpha_1,\cdots,\alpha_N}$ need not be measured.  Full QST over $\rho_l$ is then no longer required, and the complexity of evaluating  Eq.~(\ref{fid2}) or of setting a lower bound on ${\cal F}^2(\rho_l, \rho_\psi)$ can be greatly reduced.


A straightforward example of using symmetry to simplify fidelity estimation can be seen in previous work with $N$-qubit cat states $\ket{{\sf GHZ}}_N=\frac{1}{\sqrt{2}}(\ket{0_10_2\cdots 0_N}+\ket{1_11_2\cdots 1_N})$ in trapped ion systems~\cite{sackett00,leibfried04}.  The $\ket{{\sf GHZ}}_N$ state is uniquely defined by the symmetry operators $\{ \sigma_x^1  \sigma_x^2   \cdots   \sigma_x^N , \sigma_z^1 \sigma_z^2,\sigma_z^2   \sigma_z^3, \cdots, \sigma_z^{N-1}  \sigma_z^N\}$, that leave the state unchanged after their action.  As we will show, the fidelity of having prepared $\ket{{\sf GHZ}}_N$ can be estimated by measuring the expectation values of the symmetry operators. In an ion-trap setup, for example, repeated simultaneous measurements of the projections of all of the the ion spins along the $x$ axis, and of all of the ion spins along the $z$ axis, gives the fidelity of having prepared the $\ket{{\sf GHZ}}_N$ state~\cite{ion}.

In Sec.~\ref{simstate} we expand this idea to study certain cases in which the desired state can be characterized by different types of symmetries.  First, we focus on the class of rotational-invariant states (i.e., eigenstates of the total angular momentum operator) since some interesting entangled states for quantum information tasks are in this class~\cite{bre05}. Second, we study the family of stabilizer states (SSs) which  provide the foundation of the stabilizer formalism used in different quantum error-correcting procedures~\cite{gott97}.  Third, we study the case of generalized coherent states (GCSs) which provide a natural framework to study certain quantum simulations of many-body problems~\cite{som04,som06}.
In Sec.~\ref{trappedions} we apply the obtained results to estimate (numerically) the fidelity of evolving the internal states of two trapped ions with an Ising-like Hamiltonian, using the methods described in Ref.~\cite{por04}.
Finally, in Sec.~\ref{error} we discuss the estimation of the fidelity of state preparation due to the statistics from a finite number of experiments, and in Sec.~\ref{concl} we present the conclusions.

\section{Quantum fidelity and highly symmetric states}
\label{simstate}

The density operator of
a pure state $\ket{\psi}$, uniquely characterized by its symmetry operators $\{\hO_1, \cdots, \hO_L \}$,
can be written in terms of these operators only.
Thus,  the fidelity of having prepared $\ket{\psi}$ [Eq.~(\ref{fid1})] can be estimated by measuring
observables, over the actual prepared state $\rho_l$, that solely involve correlations between the $\hO_k$'s. In other words,  measurements in bases not related to the symmetry operators are not required because they do not provide any information when evaluating the fidelity of state preparation. The purpose of this section is  then to give lower bounds for estimating the fidelity  of state preparation for  three classes of  highly-symmetric $N$-qubit quantum states, and show that these can be efficiently obtained.

\subsection{Rotational-invariant states}

For a system of $N$ qubits, the rotational-invariant pure states are completely specified by the equations
\begin{eqnarray}
\label{angmom}
J^2 \ket{\psi} = j(j+2) \ket{j,j_z}, \\
\label{jz}
J_z \ket{\psi} = j_z \ket{j,j_z},
\end{eqnarray}
where $J^2= J_x^2 + J_y^2 + J_z^2$ is the (squared) total angular momentum operator, $J_\gamma = \sigma_\gamma^1 + \sigma_\gamma^2 + \cdots + \sigma_\gamma^N$ ($\gamma=x,y,z$), and $\sigma_\gamma^j$ is the corresponding Pauli operator acting on the $j$th qubit.  The factor 2 in Eq.~(\ref{angmom}) is because we are using Pauli operators instead of the actual spin-1/2 operators. Then,
the quantum numbers $j$ and $j_z$ satisfy the following properties: $j^{\max}=j_z^{\max}=N$, $|\Delta j| \ge 2$, $|\Delta j_z |\ge 2$, and $-N \le j_z \le N$ (the symbol $\Delta$ indicates the difference between the corresponding eigenvalues).
In particular, if $-N+2 \le  j_z \le N-2$ the state $\ket{j,j_z}$ is entangled
 and for $j=N$, $j_z=N-2$, then $\ket{N,N-2} = \ket{W_N}$, with
\begin{eqnarray}
\nonumber
\ket{W_N}= \frac{1}{\sqrt{N}} [\ket {1_10_2 \cdots 0_N}& +& \ket{0_11_2 \cdots 0_N}  + \\
\label{wstate}
\cdots &+&\ket{0_10_2\cdots 1_N}].
\end{eqnarray}
Although the $\ket{W_N}$ states are not maximally entangled for $N>2$, they are particularly useful for processes such as teleportation~\cite{joo03}.

The density operator $\rho_{j,j_z}$ of an $N$-qubit rotational-invariant state with quantum numbers $j=N$ and $-N\le j_z \le N$, in terms of the symmetry operators $J$ and $J_z$,  is
\begin{equation}
\label{proj1}
\rho_{j,j_z}=\kappa^{-1} \left[ \hat{ \prod}_{  -j \le j'_z \le j} \   \hat{\pi}_{j'_z}  \right] \left[\hat{\prod}_{ 0 \le j' \le N} \hat{\pi}_{j'}\right],
\end{equation}
where $\hat{\pi}_{j'_z}=(J_z - j'_z)$ and
 $\hat{\pi}_{j'} =[J^2-j'(j'+2)]$. The symbol $\hat{\prod}$ denotes that
the  term $\hat{\pi}_{j,j_z}$ has been excluded from the product.
The normalization constant $\kappa$ is given by
\begin{equation}
\kappa=  \hat{\prod }_{ -j \le j'_z \le j} \  \hat{ \prod }_{ 0 \le j' \le N} (j_z - j'_z)[j(j+2)- j'(j'+2)].
\end{equation}
To evaluate the fidelity of Eq.~(\ref{fid1}), that is  ${\cal F}(\rho_l,\rho_{j,j_z})=[{\sf Tr}(\rho_l \rho_{j,j_z})]^{1/2}$, it suffices to obtain the expectations of the correlations between the operators $J^2$ and $J_z$ appearing in Eq.~(\ref{proj1}) only. Although this procedure is still inefficient and an exponentially large number (with respect to $N$) of observables (i.e., products of Pauli operators) must be measured, it is more resource-efficient than performing full QST to obtain ${\cal F}(\rho_l,\rho_{j,j_z})$.

For example, if one is interested in preparing the Bell state $\ket{{\sf {Bell}}}=\ket{j=2,j_z=0}=\frac{1}{\sqrt{2}}[\ket{1_10_2}+\ket{0_11_2}]$ on  an ion-trap device, the fidelity of faithful preparation could be obtained by performing measurements over three different bases only, corresponding to the expectations $\langle \sigma_x^1  \sigma_x^2 \rangle_{\rho_l}$, $\langle \sigma_y^1  \sigma_y^2 \rangle_{\rho_l}$,
$\langle \sigma_z^1   \rangle_{\rho_l}$, $\langle  \sigma_z^2 \rangle_{\rho_l}$,
and $\langle \sigma_z^1  \sigma_z^2 \rangle_{\rho_l}$, respectively.  

To obtain
a lower bound on the fidelity of rotational-invariant state preparation, for $j=N$, we first define the operators ${\cal S}_{J_z} = -\frac{1}{4}(J_z -j_z)^2$ and ${\cal S}_{J^2}= -\frac{1}{64}(J^2 -N(N+2))$. 
These satisfy
\begin{equation}
\label{proper1}
\left[ {\cal S}_{J_z} + {\cal S}_{J^2} \right]   \ket{j',j'_z}  =  e_{j',j'_z} \ket{j',j'_z},
\end{equation}
with $e_{j',j'_z} \le -1$  for $(j',j'_z) \ne (j,j_z)$ and $e_{j,j_z}=0$.
Therefore, for a general pure state
 $\ket{\phi}= \sum_{j',j'_z} c_{j',j'_z} \ket{j',j'_z}$, we obtain
\begin{equation}
\label{lbound1}
\bra{\phi} {\cal S}_{J_z} + {\cal S}_{J^2} +1\ket{\phi} =  \sum_{j',j'_z}  (e_{j',j'_z}+1) |c_{j',j'_z}^2 |
\le  |c_{j,j_z}|^2,
\end{equation}
where $|c_{j,j_z}|^2$ is the probability of projecting $\ket{\phi}$ onto the state $\ket{j=N,j_z}$ (i.e., the squared fidelity between the states).
Since the actual prepared state $\rho_l$ is in general a convex combination of pure states, Eq.~(\ref{lbound1}) yields to
\begin{equation}
{\cal F}^2(\rho_l, \rho_{j,j_z} )\ge \langle {\cal S}_{J_z} + {\cal S}_{J^2} \rangle_{\rho_l} +1.
\end{equation}
This lower bound can be efficiently estimated by measuring only the observables $J_z$,  $J_z^2$, and $J^2$ a large number of times over the state $\rho_l$.  This corresponds to the measurement of $3N^2-2N$ expectations of different products of Pauli operators; that is, polynomial in $N$.

When $j<N$, the subspace with quantum numbers $j,j_z$ is degenerate and Eqs.~(\ref{angmom}) and~(\ref{jz}) do not specify the state uniquely. Then, Eq.~(\ref{proj1}) becomes the projector onto the corresponding subspace. Nevertheless, the squared fidelity ${\cal F}^2(\rho_l,\rho_{j,j_z})$ will still denote the probability of having created a pure or mixed quantum rotational-invariant state with quantum numbers $j,j_z$. When $j<N$, the operator ${\cal S}_{J^2}$ must be redefined as ${\cal S}_{J^2}= -\frac{1}{64}(J^2 -j(j+2))^2$, so the properties for the coefficients $e_{j',j'_z}$ in Eq.~(\ref{proper1}) still hold. In this case, a lower bound to ${\cal F}^2(\rho_l,\rho_{j,j_z})$ can be obtained by measuring
$N^4/4 -N^3 +11N^2/4 -N$ ($N$ even)
expectations of different products of Pauli operators.

\subsection{Stabilizer states}

Another interesting family of states are the stabilizer states~\cite{gott97}, which are defined by
\begin{equation}
\label{stab1}
\hO_s \ket{\psi} = +1 \ket{\psi} \ ; s \in[1,S] \ .
\end{equation}
The stabilizer operators $\hO_s\in \fu(2^N)$ are products of Pauli operators~\cite{stab} and have $\pm 1$ as possible eigenvalues.  (Note that $ \hO_1 = \one$ is the trivial  stabilzer.) An immediate consequence of Eq.~(\ref{stab1}) is that the operators $\hO_s$ commute with each other:  $[\hO_s, \hO_{s'}]=0$.  Here, we focus on the case when the state $\ket{\psi}$ is uniquely defined by
Eq.~(\ref{stab1}); that is,  the dimension of the stabilized space is one. The set $G_S=\{ \hO_1, \cdots, \hO_S \}$ forms the so called stabilizer group for $\ket{\psi}$. For practical purposes, we  define $G_S$ in a compact way by its $L$ linear independent generators~\cite{gott97}:  $G_S \equiv (\hg_1, \cdots, \hg_L)$, satisfying
\begin{equation}
\hg_i \ket{\psi} = +1 \ket{\psi} \ ; \ i\in [1,L].
\end{equation}
Without loss of generality we can write $\ket{\psi}\equiv\ket{g_1=1,\cdots,g_L=1}$.

The eigenstates of the stabilizer operators (associated with the stabilizer state) form a complete set of the $2^N$ dimensional Hilbert space ${\cal H}$. Therefore, the  density operator $\rho_\psi$ can be written within this formalism as ($\one\equiv \one^1 \otimes \cdots \otimes \one^N$):
\begin{eqnarray}
\nonumber
\rho_\psi&=&\ket{g_1=1,\cdots,g_L=1} \bra{g_1=1,\cdots,g_L=1} \\
\label{dens-stab}
&=& \frac{1}{2^L} \prod_{i=1}^L (\hg_i +\one),
\end{eqnarray}
and the fidelity  [Eq.~(\ref{fid1})] can be estimated by measuring, over the actual state $\rho_l$,
 the expectations of operators appearing in Eq.~(\ref{dens-stab}).

A  lower bound on the fidelity can be obtained in this case by defining the operator  ${\cal S} _{G_S}= \frac{1}{2}[(\sum_{i=1}^L \hg_i )- (L-2)\one]$. Then,
\begin{equation}
{\cal S}_{G_S} \ket{g_1,\cdots,g_L} = e_{g_1,\cdots,g_L} \ket{g_1,\cdots,g_L} , \ (g_i=\pm1),
\end{equation}
with $e_{1,\cdots,1}=1$ and $ e_{g_1,\cdots,g_L}  \le 0$ otherwise. Following the same procedure used for rotational-invariant states, we arrive to the inequality
 \begin{equation}
 \label{lbound2}
{\cal F}^2 (\rho_l,\rho_\psi)\ge \langle {\cal S}_{G_S} \rangle_{\rho_l},
 \end{equation}
which  can be efficiently estimated by measuring the expectations $\langle \hg_i \rangle_{\rho_l} \ \forall i\in [1,L]$.

As an example we consider the the Bell state $\ket{\sf Bell}=\frac{1}{\sqrt{2}} [\ket{0_11_2}-\ket{1_10_2}]$.  For this state, the stabilizer group is defined by the generators $G_S \equiv (-\sigma_z^1  \sigma_z^2, -\sigma_x^1 \sigma_x^2)$. Then, $L = 2$ and ${\cal S} _{G_S}= \frac{1}{2}[-\sigma_z^1  \sigma_z^2 -\sigma_x^1 \sigma_x^2]$. Another example is the set of maximally entangled $N$-qubit states $\ket{{\sf GHZ}}_N = \frac{1}{\sqrt{2}}[\ket{0_1 0_2 \cdots 0_N} + \ket{1_1 1_2 \cdots 1_N}]$.  For these states, the generators of the corresponding stabilizer group are given by $G_S \equiv (\sigma_x^1 \sigma_x^2  \cdots  \sigma_x^N, \sigma_z^1  \sigma_z^2, \sigma_z^2   \sigma_z^3, \cdots, \sigma_z^{N-1} \sigma_z^N )$, as pointed out in Sec.~\ref{intro}.  For $N = 3$, $L = 3$ and ${\cal S} _{G_S}= \frac{1}{2}[\sigma_x^1  \sigma_x^2 \sigma_x^3 + \sigma_z^1  \sigma_z^2 + \sigma_z^2  \sigma_z^3 - \one]$.

\subsection{Generalized coherent states}
\label{GCSSection}

The last class of states we consider are the generalized coherent states (GCSs)~\cite{zha90}. For a semi-simple, compact, $M$-dimensional Lie algebra $\fh=\{\hQ_1,\hQ_2,\cdots,\hQ_M\}$, with $\hQ_j=(\hQ_j)^\dagger$ the $N$-qubit operators acting on the $2^N$ dimensional Hilbert space ${\cal H}$, the GCSs are defined via
\begin{equation}
\label{gcs}
\ket{{\sf GCS}} \equiv e^{i\fh} \ket{\sf hw}.
\end{equation}
Here, $e^{i\fh}$ denotes a unitary group operation (displacement) induced by $\fh$: $e^{i\fh} \equiv \exp[i(\sum_j \lambda_j \hQ_j)], \ \lambda_j \in \mathbb{R}$. The state $\ket{{\sf hw}}$ is  the highest-weight state of $\fh$. To define it, one needs to assume a Cartan-Weyl (CW) decomposition  $\fh=\fh_D \oplus \fh^+ \oplus \fh^-$~\cite{fuc92,cor89}. The set $\fh_D=\{\hh_1, \cdots, \hh_r \}$ is the Cartan subalgebra of $\fh$ (CSA) constructed from the largest set of commuting operators (observables) in $\fh$. The weight states $\ket{\phi_i}$, which form a  basis of states for ${\cal H}$, are  the eigenstates of $\fh_D$:
\begin{equation}
\label{weight}
\hh_k \ket{\phi_i} = u_k^i \ket{\phi_i} , \ k\in [1,r], \ i \in [0,2^N-1] \ .
\end{equation}
The sets $\fh^+=\{\he^+_{\alpha_1},\cdots, \he^+_{\alpha_l}\}$ and $\fh^-=\{\he^-_{\alpha_1},\cdots, \he^-_{\alpha_l}\}$ are built from raising and lowering operators ($\he^+_{\alpha_j}=\he_{\alpha_j}^{-\dagger}$), and either map weight states into orthogonal weight states or annihilate them. (The subscripts $\alpha_j \in \mathbb{R}^r$ are the roots of $\fh$ and are considered to be positive.) Then, $\ket{{\sf hw}}$ is defined by
\begin{eqnarray}
\label{hwdefined}
\hh_k \ket{{\sf hw}} &=& v_k \ket{{\sf hw}}, \ k \in [1,r], \\
\he^+_{\alpha_j} \ket{{\sf hw}} &=&0, \ j \in [1,l] ,
\end{eqnarray}
with $v_k = u_k^0$ (i.e., we have assumed $\ket{{\sf hw}} \equiv \ket{\phi_0}$).  Note that $M=r+2l$.  In many cases, $\ket{{\sf hw}}=\ket{0_1 0_2 \cdots 0_N}$, where $\ket{0_i}$ represents an eigenstate of $\sigma_z^i$.

As shown in Refs.~\cite{bar03,som04,som06},
when the dimension of $\fh$ satisfies $M \le \text{poly}(N)$,
the corresponding GCSs
play a decisive role in the theory of entanglement and quantum and classical simulations of many-body systems.
An example is given by the GCSs defined via
\begin{equation}
\ket{\psi_I(t)}= e^{-iH_It} \ket{0_1 0_2 \cdots 0_N},
\end{equation}
where $H_I$ is the Hamiltonian corresponding to the exactly-solvable one-dimensional anisotropic Ising model in a transverse magnetic field and periodic boundary conditions:
\begin{equation}
\label{hising}
H_I= \sum_{j=1}^N [ \gamma_x \sigma_x^j \sigma_x^{j+1} + \gamma_y \sigma_y^j \sigma_y^{j+1} + B \sigma_z^j ] \ .
\end{equation}
In section \ref{trappedions} we will discuss this system in more detail.

Any GCS is uniquely determined (up to a global phase) by the expectation values of the operators in $\fh$. The state $\ket{{\sf hw}(t)} = e^{-iHt} \ket{\sf hw}$, with $H \in \fh$, is the highest-weight state of $\fh$ in a rotated CW basis, and satisfies
\begin{equation}
\hh_k(t) \ket{{\sf hw}(t)}= v_k \ket{{\sf hw}(t)} , \ k \in [1,r],
\end{equation}
where $\hh_k(t) = e^{-iHt} \hh_k e^{iHt}=\hh_k+ i[\hh_k,H]+\cdots \in \fh$. Thus,
\begin{equation}
\label{proj-gcs}
\rho_{{\sf hw}}(t)=\ket{{\sf hw}(t)}\bra{{\sf hw}(t)}= \kappa^{-1} \prod_{k, i \ne 0} (\hh_k(t) -u_k^i \one),
\end{equation}
where  $\kappa=\prod_{k,i\ne0} (v_k-u_k^i)$ is a constant for normalization purposes. For a particular value of $t$, the operators $\hh_k(t)=\sum_{j=1}^M \lambda_j(t) \hQ_j$ can be obtained on a classical computer [i.e., the coefficients $\lambda_j(t)$]  in time polynomial in $M$  (see Theorem 1 in Ref.~\cite{som06}). To see this, note first that $\lambda_j(t) \propto {\sf Tr} [\hh_k(t) \hQ_j]$. Such a trace can be efficiently evaluated by working in the $(M \times M)$-dimensional matrix representation (or any other faithful representation) of $\fh$ rather than working in the $(2^N \times 2^N)$-dimensional original representation.
Therefore, the fidelity of having prepared $\ket{{\sf hw}(t)}$ can be obtained by measuring the expectations of the observables appearing in Eq.~(\ref{proj-gcs}), over the actual prepared state $\rho_l$.

In analogy to the previously discussed cases, a lower bound for the fidelity can be obtained  by defining the operator ${\cal S}_{\fh_D}(t) =[-\varepsilon ( \sum_k \hh_k(t) - v_k \one )^2]+\one$, with $\varepsilon > 0$  a constant determined by the spacing between the eigenvalues $u_k^i$ (see below). If $u_k^i < v_k \ \forall i \in [1,2^N-1]$ one can consider ${\cal S}_{\fh_D}(t) =[-\varepsilon ( \sum_k \hh_k(t) - v_k \one )]+\one$, instead.
Then,
\begin{equation}
{\cal S}_{\fh_D}(t) \ket{\phi_i(t)} =   w_i \ket{\phi_i(t)},
\end{equation}
where $ \ket{\phi_i(t)}=e^{-iHt} \ket{\phi_i}$ are the weight states in the rotated CW basis (e.g., $\ket{{\sf hw}(t)} \equiv \ket{\phi_0(t)}$), $w_i \in \mathbb{R}$, and $w_0=1$.
Thus,  $\varepsilon$ is chosen such that  $w_i=[-\varepsilon ( \sum_k u_k^i - v_k )^2+1] $ (when ${\cal S}_{\fh_D}(t) =[-\varepsilon ( \sum_k \hh_k(t) - v_k \one )^2]+\one$), or $w_i=[-\varepsilon ( \sum_k u_k^i - v_k )+1] $ (when ${\cal S}_{\fh_D}(t) =[-\varepsilon ( \sum_k \hh_k(t) - v_k \one )]+\one$), satisfies
\begin{equation}
\label{wsign}
w_i \le 0 \ \forall i \ne 0.
\end{equation}
For a particular value of $t$,  Eq.~(\ref{wsign})  yields
\begin{equation}
\label{lbound3}
{\cal F}^2_{\rho_l,\rho_{\sf hw}(t)}  \ge \langle  {\cal S}_{\fh_D}(t) \rangle_{\rho_l}.
\end{equation}
This lower bound
can be obtained experimentally  by measuring the expectation values of the operators $\hh_k(t) \hh_{k'}(t)$ and (or) $\hh_k(t)$, which are directly induced from the expectations
$\langle \hQ_j \rangle_{\rho_l}$  and $\langle \hQ_j \hQ_{j'}\rangle_{\rho_l} \ \forall j,j'\in[1,M]$ (assumed to be measurable with our quantum device).
If $M = \text{poly}(N)$ (e.g., an evolution due to the Ising Hamiltonian $H_I$), Eq.~(\ref{lbound3}) can be efficiently estimated with ${\cal O}[\text{poly}(N)]$ measurements.

\section{Quantum simulations with two trapped ions}
\label{trappedions}

In this section we use some of the results obtained in Sec.~\ref{GCSSection} to estimate the fidelity of evolving two trapped ions (qubits) with the Ising-like interaction
\begin{equation}
\label{hising2}
H_I=J\sigma_x^1 \sigma_x^2 + B (\sigma_z^1+ \sigma_z^2),
\end{equation}
where $J$ is the spin-spin coupling and $B$ is a transverse magnetic field. To do so, we will model the system of two ions confined in a linear Paul trap and interacting with resonant and non-resonant laser fields as described in Ref.~\cite{por04}.  We will estimate the reliability of having prepared the state $\ket{\psi(t)}= \ket{{\sf hw}(t)}= e^{-i H_I t} \ket{0_1 0_2}$ (for fixed $t$).

In this case, the interaction Hamiltonian for the ions in the trap is given by
\begin{eqnarray}
\label{htrap}
H_{\sf trap}&=& H_{\sf phonon} + H_{\sf l-ion1} + H_{\sf l-ion2} + H_{\sf m} \ , \\
\nonumber
 H_{\sf phonon}&=& \omega_{\sf cm} a^\dagger_{\sf cm} a^{\;}_{\sf cm} + \omega_{\sf br} a^\dagger_{\sf br} a^{\;}_{\sf br} \ , \\
 \nonumber
 H_{\sf l-ion1}&=&
 -[ \eta_{\sf cm}  \omega_{\sf cm} (a^\dagger_{\sf cm}+ a^{\;}_{\sf cm}) +
  \eta_{\sf br}  \omega_{\sf br} (a^\dagger_{\sf br}+ a^{\;}_{\sf br}) ] \sigma_x^1 \ ,\\
  \nonumber
  H_{\sf l-ion2}&=&
 -[ \eta_{\sf cm}  \omega_{\sf cm} (a^\dagger_{\sf cm}+ a^{\;}_{\sf cm}) -
  \eta_{\sf br}  \omega_{\sf br} (a^\dagger_{\sf br}+ a^{\;}_{\sf br}) ] \sigma_x^2  \  ,\\
  \nonumber
  H_{\sf m}&=& B (\sigma_z^1+ \sigma_z^2).
\end{eqnarray}
Here, the operators $a^\dagger_{\sf cm}$ ($a^{\;}_{\sf cm}$) and  $a^\dagger_{\sf br}$ ($a^{\;}_{\sf br}$) create (annihilate) an excitacion in the center of mass and breathing modes, respectively. The coupling interactions $H_{\sf l-ion1}$ and  $H_{\sf l-ion2}$  are due to the action of state-dependent dipole forces, which are generated by  the interaction of non-resonant laser beams with the electronic levels of the ions (see Ref.~\cite{por04}).
$H_{\sf m}$ is due to the action of an effective magnetic field that can be external or generated by resonant laser beams. $H_{\sf phonon}$ is the energy of the normal modes with frequency $\omega_{\sf cm}/2 \pi$ for the center of mass mode, and $\omega_{\sf br}/2 \pi$ for the breathing mode. In the case of a single well potential in one dimension, $\omega_{\sf br}= \sqrt{3} \omega_{\sf cm}$. The couplings (displacements) $\eta_{\sf cm}$ and $\eta_{\sf br}$ are assumed to be small: $\eta_{\sf i} \ll 1$. They depend on the intensities of the laser beams and are given by
\begin{equation}
\eta_{\sf i} = \frac{F}{\sqrt{2} \hbar \omega_{\sf i}} \sqrt{\frac{\hbar}{2m\omega_{\sf i}}},
\end{equation}
with ${\sf i}=[{\sf cm, br}]$, $F$  the dipole force acting on each ion, and $m$ the mass of the ion.

Therefore, for a fixed value of $t$, the actual two-qubit state prepared in the ion-trap device is
\begin{equation}
\rho_l(t)= {\sf Tr_{phonon}}[e^{-iH_{\sf trap}t}\rho_{\sf (ion-phonon)} e^{iH_{\sf trap}t}],
\end{equation}
where we have traced out the vibrational modes.
Here, the initial state is $\rho_{\sf (ion-phonon)}= \ket{0_1 0_2}\bra{0_1 0_2} \otimes \rho_{\sf phonon}$, and
$\rho_{\sf phonon} \propto e^{-\frac{ H_{\sf phonon}}{KT}}$ is the density operator for the initial state of the phonons, with the ion motion in a thermal distribution being at temperature $T$ ($K$ is the Boltzmann constant).
The fidelity of having prepared the state $\ket{{\sf hw}(t)}$ is then given by
\begin{equation}
\label{fidtrap}
{\cal F}^2({\rho_l(t),\rho_{\sf hw}(t)} )= {\sf Tr} [\rho_l(t) \rho_{\sf hw}(t)],
\end{equation}
where the trace is over the spin (i.e., two-qubit) degrees of freedom.

Following the results of Sec.~\ref{GCSSection}, we first identify the set $\fh_D = \{\sigma_z^1, \sigma_z^2\}$ as the largest set of commuting observables in $\fh$.  This determines $\ket{\sf hw} = \ket{0_1  0_2}$ according to Eq.~(\ref{hwdefined}).
A bound for the fidelity of Eq.~(\ref{fidtrap}) can be obtained by using the time dependent symmetry operators
\begin{equation}
\tilde{\sigma}_z^j(t)=e^{-iH_I t} \sigma_z^j e^{iH_I t} \ \ (j=1,2),
\end{equation}
that uniquely define the state $\ket{{\sf hw}(t)}$ through the equations
\begin{equation}
\tilde{\sigma}_z^j(t) \ket{{\sf hw}(t)}=+1\ket{{\sf hw}(t)}.
\end{equation}
Choosing $\varepsilon=1/2$ (see Sec.~\ref{simstate}) and considering that $v_1=v_2=1$, we obtain
${\cal S}_{\fh_D}(t)= \frac{1}{2}[\tilde{\sigma}_z^1(t)+  \tilde{\sigma}_z^2(t)]$, which satisfies [Eq.~(\ref{lbound3})]
\begin{equation}
\label{lbound7}
{\cal F}^2({\rho_l(t),\rho_{\sf hw}(t)}) \ge \langle {\cal S}_{\fh_D}(t) \rangle_{\rho_l(t)}.
\end{equation}

The  $\tilde{\sigma}_z^j(t)=(\sigma_z^j-it[H_I,\sigma_z^j]+ \cdots)$ are linear combinations of operators belonging to
the Lie algebra $\fso(4)=\{\sigma_z^1, \sigma_z^2, \sigma_x^1\sigma_x^2, \sigma_x^1 \sigma_y^2, \sigma_y^1 \sigma_x^2, \sigma_y^1 \sigma_y^2 \}$. To obtain the coefficients involved in these combinations one needs to find the trace between the corresponding operators. For example, to obtain the coefficient $\lambda_1(t)$ that accompanies the operator $\sigma_z^1$ in the decomposition of $\tilde{\sigma}_z^1(t)$, one needs to compute $\frac{1}{4}{\sf Tr} [\sigma_z^1 \tilde{\sigma}_z^1(t)]$. Remarkably, such a trace can be efficiently computed by working in the $(2N \times 2N)$-dimensional fundamental matrix representation of $\fso(2N)$ rather than in the $(2^N \times 2^N)$-dimensional original representation (see Ref. \cite{som06} for details).

In brief, only six correlations (i.e.,  the elements of $\fso(4)$) need to be measured to evaluate the inequality of Eq.~(\ref{lbound7}). The complexity of estimating the fidelity is then reduced since a naive approach to fidelity estimation would involve the measurement of fifteen correlations (i.e., the elements of the algebra $\fsu(4)$).
Of course, the complexity of the problem is slightly reduced in this case but the difference is much greater for larger systems.

\begin{figure}[hbt]
\begin{center}
\includegraphics*[height=11cm]{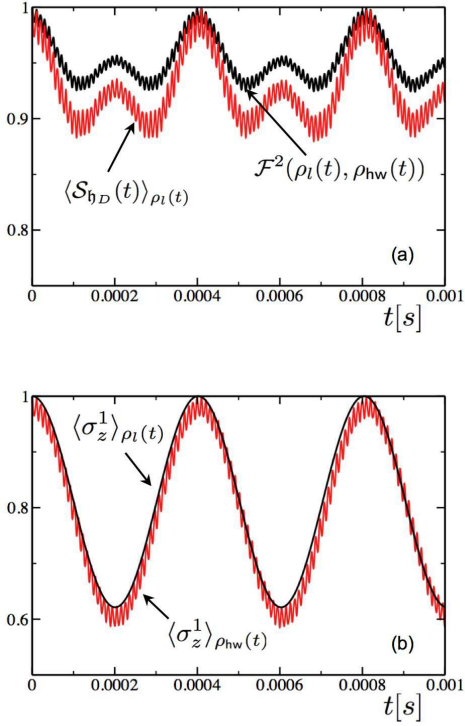}
\end{center}
\caption{Numerical simulation of the quantum evolution of two trapped ions interacting with laser fields. The parameters used are $\omega_{\sf cm}=100kHz$, $\eta_{\sf cm}\approx 0.063$, $B=560Hz$, $-J=540Hz$,  $F=25.10^{-23}N$, and $T=0$. These are expected to be attained experimentally.
(a) Squared fidelity (probability) of having prepared the state $\ket{{\sf hw}(t)}=e^{-iH_It}\ket{0_1 0_2 }$, if the dynamics of the trapped ions is dominated by the trap Hamiltonian $H_{\sf trap}$ [Eq.~(\ref{htrap})], and the corresponding lower bound $\langle {\cal S}_{\fh_D}(t) \rangle_{\rho_l(t)}$, as given by Eq.~(\ref{lbound7}), as a function of time. 
(b)  Expectations of the Pauli operator $\sigma_z^1$ as a function of time, if the evolution is governed by $H_I$ and $H_{\sf trap}$, respectively. }
\label{figbound}
\end{figure}

\begin{figure}[hbt]
\begin{center}
\includegraphics*[height=6cm]{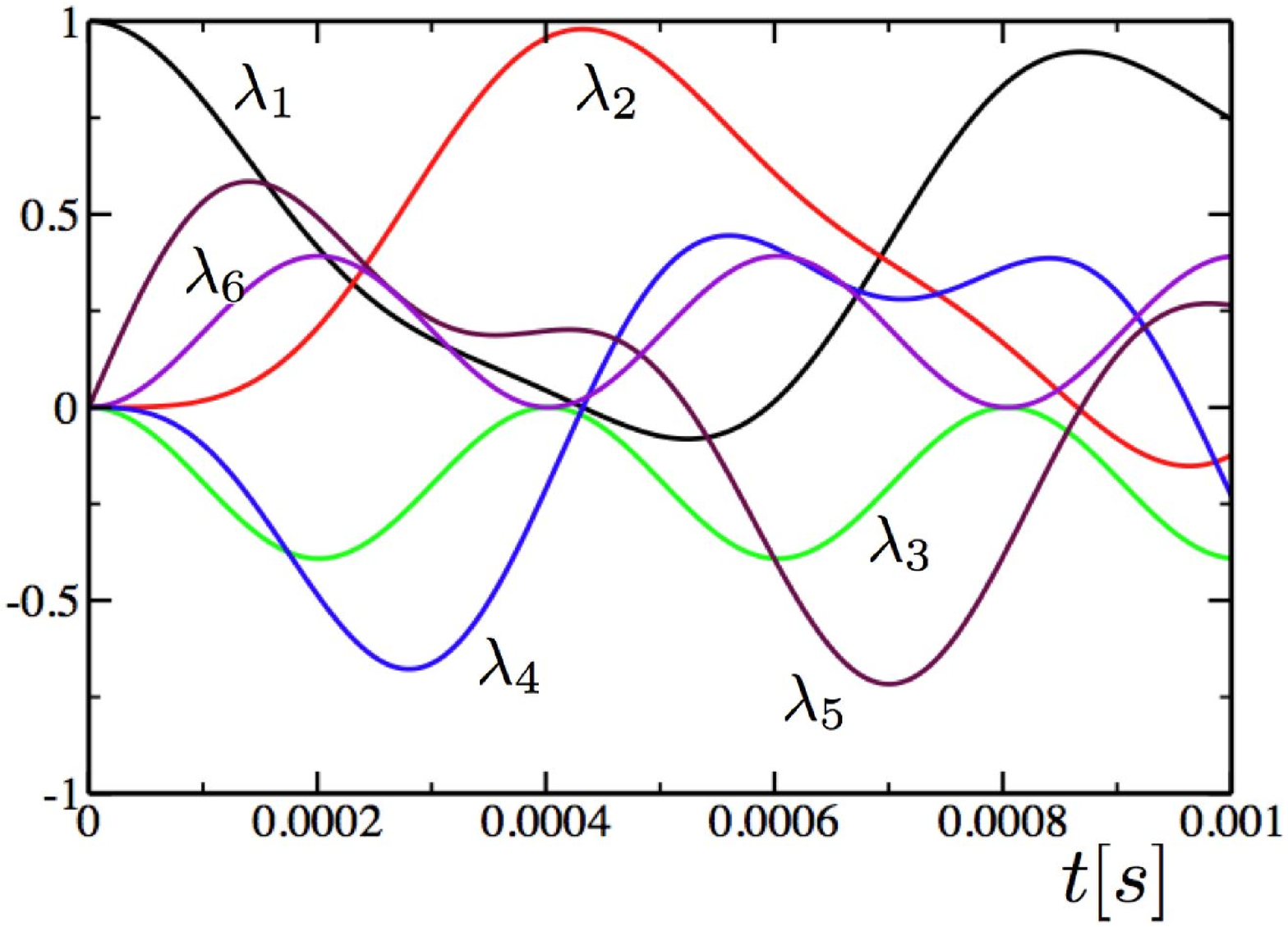}
\end{center}
\caption{Coefficients $\lambda_j(t)$, where $\tilde{\sigma}_z^1(t)=\lambda_1^1(t)\sigma_z^1 + 
\lambda_1^2(t)\sigma_z^2 + \lambda_1^3(t)\sigma_x^1 \sigma_x^2+ \lambda_1^4(t)\sigma_x^1\sigma_y^1 + \lambda_1^5(t)\sigma_y^1 \sigma_x^2+ \lambda_1^6(t)\sigma_y^1 \sigma_y^2  
\in \fso(4)$,
 used to obtain $\langle {\cal S}_{\fh_D} \rangle_{{\rho_l(t)}}$ in Fig.~\ref{figbound}. 
 Note that, because of the symmetry under permutation of both ions,  the same coefficients are obtained in the decomposition of $\tilde{\sigma}_z^2(t)$.
}
\label{figcoeff}
\end{figure}

In Fig.~\ref{figbound} we plot ${\cal F}^2(\rho_l(t),\rho_{\sf hw}(t))$ [Eq.~(\ref{fidtrap})] as a function of time and for certain values of $F$, $\omega_{\sf i}$, and $B$ that could be attained experimentally. We observe that, for these parameters, the fidelity remains close to one, implying that the ion-trap device can be used to perform a quantum simulation governed by the Ising-like Hamiltonian of Eq.~(\ref{hising2}). We also plot $\langle {\cal S}_{\fh_D}(t)\rangle_{\rho_l(t)}$ and we observe that this lower bound of the (squared) fidelity already describes much of the reliability of the simulation. For the sake of comparison,  we also plot the expectations $\langle \sigma_z^1 \rangle_{\rho_{\sf hw}(t)}$ and $\langle \sigma_z^1 \rangle_{\rho_{l}(t)}$.  Finally, in Fig.~\ref{figcoeff} we plot the coefficients $\lambda_j(t)$, $j\in[1,6]$, that determine the weighting of the six correlation measurements that contribute to the estimate of $\langle {\cal S}_{\fh_D}(t)\rangle_{\rho_l(t)}$.

\section{Statistical contributions to measured lower bound on the fidelity}
\label{error}
In an actual experiment, expectation values can never be exactly obtained due to quantum projection noise.  Thus, they must be estimated after a (typically large) sequence of projective measurements performed on
identically prepared copies of the system. Commonly, maximum-likelihood methods (MLMs)~ \cite{hradil97,banaszek00} are used to estimate the most probable density matrix $\bar{\rho}_l$ from these measurements. As with full QST, these methods are usually inefficient, and they require input data concerning every correlation in the system. For example, if a MLM is used to estimate the density operator $\rho_l$ of
an $N$-qubit system, the estimation $\bar{\Sigma}_{\rho_l}$ of the expectation of a particular operator $\Sigma= \sigma_{\alpha_1}^1   \cdots  \sigma_{\alpha_N}^N$ will require ${\cal O}[(4^N-1)X]$ identically prepared copies of $\rho_l$, where $X$ is the number of copies used to measure a particular correlation (product of Pauli operators)~\cite{ion2}. Such a complexity would then be translated to the estimation of the lower bounds of Eqs.~(\ref{lbound1}),~(\ref{lbound2}), and~(\ref{lbound3}).
In this section we argue that to estimate these lower bounds with certain (fixed) level of confidence, the exponential complexity can be avoided.

To prove this, we use results regarding the binomial distribution~\cite{taylor97}.  Observe first that the operator $\Sigma$, as defined above, has $\pm 1$ as possible eigenvalues. Then, if we perform projective measurements of $\Sigma$ over $X$ identical copies of $\rho_l$, we obtain
\begin{equation}
\label{estimate}
\langle \Sigma \rangle_{\rho_l}= \bar{\Sigma}_{\rho_l} \pm \delta,
\end{equation}
where $\bar{\Sigma}_{\rho_l}=\frac{X_+ - X_-}{X}$ is the estimated expectation (i.e., $X_\pm$ are the number of times we measured $\Sigma=\pm 1$, respectively),
 and  $\delta$ is the corresponding standard deviation. The latter is given by
\begin{equation}
\delta=2\sqrt{\frac{p_+p_-}{X}},
\end{equation}
where $p_\pm$ are the (not known) probabilities of measuring $\Sigma= \pm1$, respectively. Then, $\delta \le \sqrt{1/X}$.

For sufficiently large $X$, the binomial distribution can be well approximated by the normal distribution. In this context, Eq.~(\ref{estimate}) guarantees that  $\bar{\Sigma}_{\rho_l}$ differs by at most $\sqrt{1/X}$ from the actual expectation with (at least) $68 \%$  of confidence~\cite{chernoff}.
For example,  if $\Sigma$ is estimated from ten thousand identical copies of $\rho_l$, then $\langle \Sigma \rangle_{\rho_l} = \bar{\Sigma}_{\rho_l} \pm .01$ with (at least)  $68 \%$  of confidence.

With no loss of generality, the bounds of Eqs.~(\ref{lbound1}), (\ref{lbound2}), and (\ref{lbound3}), can be rewritten as
\begin{equation}
\label{lbound5}
{\cal F}^2 \ge a_0 + \sum_{m=1}^R a_m \langle \Sigma^m \rangle_{\rho_l}, \ a_0,  a_m\in \mathbb{R},
\end{equation}
where each $\Sigma^m$ involves a particular product of Pauli operators [$R=\text{poly}(N)$]. If each
$ \langle \Sigma^m \rangle_{\rho_l}$ is estimated from $X$ identical copies of $\rho_l$, then
$ \langle \Sigma^m \rangle_{\rho_l}=\bar{\Sigma}^m_{\rho_l} \pm \sqrt{1/X}$ with $68 \%$  of confidence, and
\begin{equation}
\label{lbound4}
{\cal F}^2 \ge a_0 + \sum_{m=1}^R a_m \langle \Sigma^m \rangle_{\rho_l} \ge a_0 + \sum_{m=1}^R a_m \bar{\Sigma}^m_{\rho_l} -R/\sqrt{X},
\end{equation}
with the same confidence. Of course, Eq.~(\ref{lbound4}) provides relevant information if $\bar{\Sigma}^m_{\rho_l} \gg 1/\sqrt{X}$. For example, if one is interested in preparing the state $\ket{{\sf GHZ}_N}$, then $R=N$ and $\bar{\Sigma}^m_{\rho_l} \approx +1$. Choosing $X=10^4 N^2$, a good estimation (with error $0.01$) for the lower bound of the fidelity is obtained. The method is then efficient:   lower bounds on fidelity of state preparation can be obtained, with certain confidence, in $\text{poly}(N)$ identical preparations of $\rho_l$.

We  have not considered any source of error other than the one given by the statistics of projective measurements in the quantum world.   Otherwise, the results obtained in the previous sections must be modified according to the specific sources of error or decoherence that can affect the state preparation.

\section{Conclusions}
\label{concl}
We have studied the fidelity of state preparation for three different classes of states: the rotational-invariant states, SSs, and GCSs. Many interesting multi-partite entangled states, like cat or W-type states, belong to these classes. In particular, GCSs are natural in the framework of quantum simulations.  We have discussed the quantum simulation of the two-qubit Ising model using an ion-trap device. In this case we observe that a lower bound of the fidelity of the simulation can be simply obtained, and can be considered to estimate the reliability of the experiment. Such a bound can also be efficiently estimated for  other multiple qubit systems having Ising-like interactions. Similar approaches can be considered to study the fidelity of state preparation in  general qudit or fermionic systems.

Our results provide an efficient method to estimate, with certain confidence, lower bounds on the fidelity of state preparation based on symmetries.  Many of the states described contain $N$-particle entanglement, so the lower bounds can also be used to verify entanglement using  entanglement witnesses~\cite{lew00,bou04}.
These bounds are most accurate when the actual prepared state is not too far from the desired one, as in Fig.~\ref{figbound}. Therefore, a consequence of our results is  that instead of measuring every possible quantum correlation of a system a large number of times (as for QST), one should focus on having good estimations of certain relevant expectations.


\acknowledgments
RS is grateful to the QiSci at the University of Queensland, G. Vidal, J. Harrington, F.M. Cucchietti, S. Boixo, and M. Boshier for discussions. This work has been supported by LANL Laboratory Directed Research and Development (LDRD).


\begin{thebibliography}{10}
\bibitem{stab}
One can work in a rotated basis where the stabilizers are linear combinations of products of Pauli operators.
\bibitem{ion2}
It is ${\cal O}(3^NX)$ experiments for an ion-trap device.
\bibitem{chernoff}
If $X$ is not large enough, one can use the Chernoff's bound to estimate the confidence.
\end{thebibliography}
\end{document}